**Tunable Extremely Asymmetric Acoustic Metasurfaces Made by Coupled Membrane Resonators**

Songwen Xiao, Suet To Tang, and Z. Yang*

[1]*Department of Physics, Hong Kong University of Science and Technology*

*Clear Water Bay, Kowloon, Hong Kong, China*

## Abstract

We report the experimental demonstration of tunable asymmetric acoustic metasurfaces with extreme contrast, made by two decorated membrane resonators (DMRs) coupled via a sealed air column. The front side of the metasurface is impedance matched to air and perfectly absorbing, while the backside is hard and totally reflecting. When a suitable DC voltage is applied to the backside DMR with proper electrodes, the surface impedance of the back side of the metasurfaces can be tuned from one extreme to the other, such that the backside becomes impedance matched to air and totally absorbing, while the front side becomes perfectly reflecting. The metasurface also exhibits high transmission contrast around two frequencies. The tunability of the reflection is over 23 dB at 388 Hz, and that of the transmission is over 33 dB at 240 Hz and 590 Hz with 600 V of applied voltage. We further demonstrate one-side impedance matched metasurface with tunable working frequency 324.2 Hz to 335 Hz and reflection contrast over 31 dB.

Acoustic wave absorption has been intensively studied in the past decades due to its great importance in both scientific research and engineering applications. For conventional materials such as porous material [1], or micro-perforated plate absorber [2], large thickness is usually required in order to maintain their acoustical performance at low frequency, which definitely limits their usage in real applications. In the past decade, substantial attention has been focused onto the local resonators based acoustic metamaterials [3, 4] to overcome the challenges faced by the conventional materials. Dissipation energy density could be greatly increased through resonant states, and with careful design of the structure parameters to balance dissipation and scattering effects, perfect absorption have been experimentally achieved with membrane resonators [5 – 7] that is sub-wavelength in physical size. Structures that combined space-coiling and Helmholtz resonators could also achieve high absorption [8 – 10]. The concept of coherent perfect absorber originally developed for electromagnetic waves [11 – 13] that relied on two coherent counter-propagating waved with specific phase and amplitude to achieve perfect absorption have been demonstrated recently with acoustic waves as well [14]. For an acoustic metasurface, no contrast would be more extreme than being perfectly absorbing or totally reflecting. Extremely asymmetric absorbers that have one surface being perfectly absorbing and the other surface totally reflecting have also been experimentally demonstrated in waveguide flanked either by a pair of Helmholtz resonators [15] or by a pair of hybrid membrane resonators [16]. These asymmetric absorbers were mounted on the sidewall of a waveguide. Their configuration makes it difficult to turn the whole area of acoustic barriers into perfectly absorbing.

In this paper we report the experimental demonstration of a coupled membrane resonator (CMR) that exhibits extremely asymmetric properties of its two metasurfaces at low frequency, i. e., the front surface is perfectly absorbing while the back surface is totally reflecting. The configuration makes it straightforward to assemble into planar sound barriers with the entire surface being perfectly absorbing. We further demonstrate that the two surfaces of the CMR can be tuned by applied DC voltage, such that the front surface becomes totally reflection while the back surface becomes perfectly absorbing. At other frequencies, the transmission of the metasurface can be tuned by over 30 dB with the applied voltage. Lastly, we experimentally demonstrate that the frequency at which perfect absorption occurs can be tuned by applied voltage by over 10.8 Hz for a one-sided metasurface with a hard wall back surface.

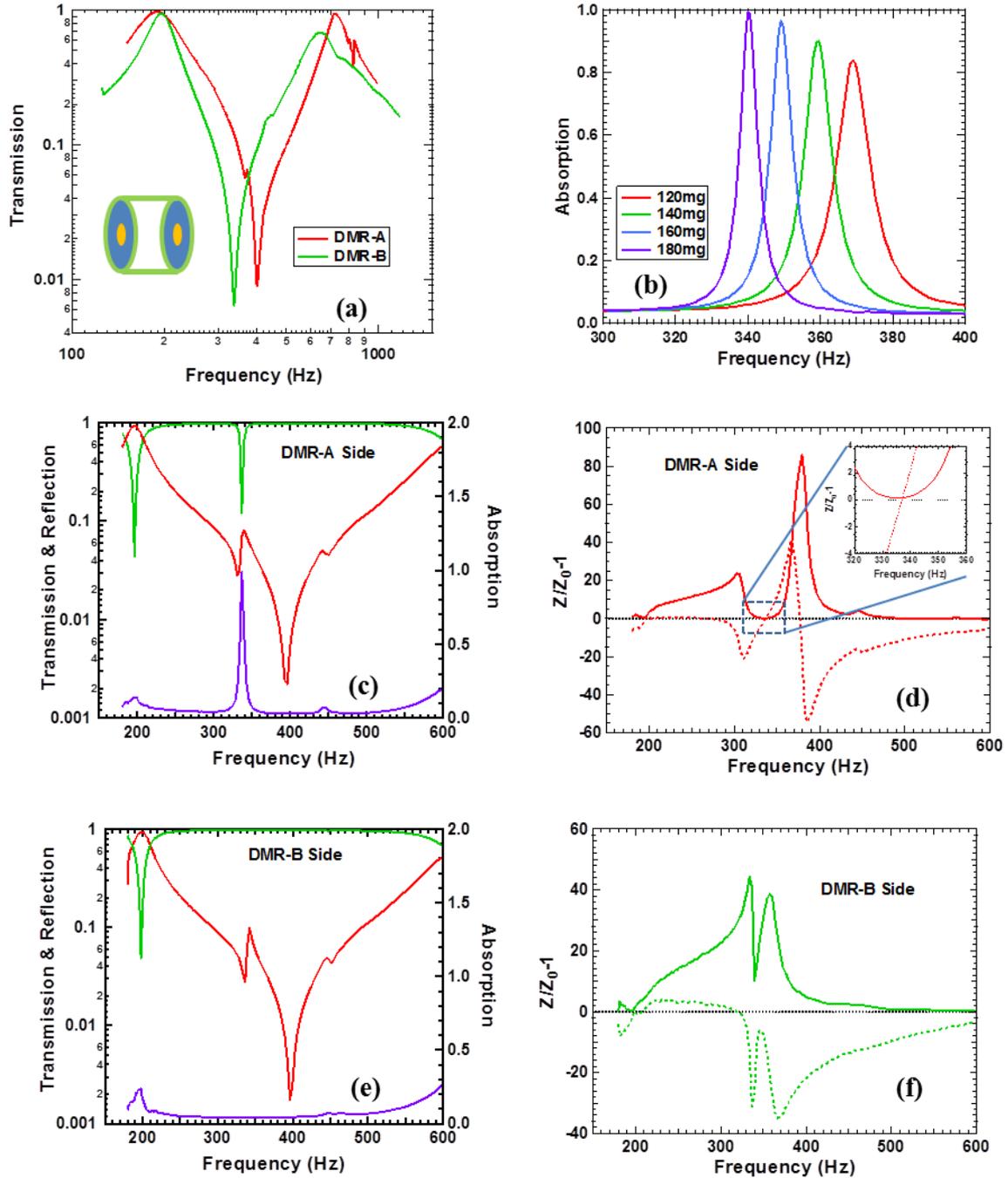

Figure 1. (a) The transmission spectra of DMR-A and DMR-B when measured individually. The insert is a schematics of the device. (b) The absorption spectra of the device at various disk mass of DMR-A when sound waves were incident to the DMR-A side. (c) The transmission (red curve, left axis), reflection (green curve, left axis), and absorption (purple curve, right axis) of the device when sound waves were incident to the DMR-A side. (d) The corresponding surface impedance of the DMR-A side of the device. The solid curve is for the real part of the impedance, while the dashed curve is for the imaginary part. (e) The transmission (red curve, left axis), reflection (green curve, left axis), and absorption (purple curve, right axis) of the device when sound waves were incident to the DMR-B side. (f) The corresponding surface impedance of the DMR-B side of the device. The solid curve is for the real part of the impedance, while the dashed curve is for the imaginary part.

The structure of the CMR is schematically shown in the insert of Fig.1a. It consisted of two decorated membrane resonators (DMRs), denoted as DMR-A and DMR-B, respectively, mounted on a rigid hollow cylinder of 25 mm in length. The sealed space between the DMRs was filled with air. The two DMRs were therefore coupled via the air column in between. DMR-A consisted of a circular rubber membrane 27 mm in radius mounted on a rigid frame and a central disk 9 mm in radius and its mass was adjusted from 120 to 180 mg. The change of mass was accomplished by adding/removing putty stuck to the disk. DMR-B consisted of a membrane of the same radius as DMR-A, and a central disk of 6 mm in radius and 160 mg in mass. The thickness of the membrane is about 0.15 mm for both DMRs. More pre-stress was intentionally introduced in the membrane of DMR-A than that of DMR-B, so that the first two eigen-frequencies of DMR-A are higher than that of DMR-B. A modified impedance tube method [17] was employed to conduct the acoustic measurements. The inner radius of the impedance tube was 50 mm. The sample was mounted on a rigid aluminum plate with a hole of proper size, and the whole assembly could fill up the inner cross section of the impedance tube.

Figure 1(a) shows the transmission spectra of the two DMRs when they were measured individually. The disk mass of DMR-A was the optimum value of 180 mg. Both spectra show typical two-peak and one-dip feature of a simple DMR [3] within the frequency range below 1500 Hz. The first transmission peak slightly below 200 Hz is due to the first eigenmode of the DMRs. They happen to be nearly the same, even though the disk mass of DMR-A is heavier than that of DMR-B, because both the tension in the membrane and the radius of the disk in DMR-A are larger. The second transmission peak around 720 Hz is due to the second eigenmode of DMR-A. It is higher than that of DMR-B at 640 Hz because of the higher tension in the membrane, and because the area not in direct contact with the central disk is smaller. As a result, the transmission dip frequency due to the anti-resonance of DMR-A is around 400 Hz, while that of DMR-B is around 340 Hz.

According to the hybrid resonance mechanism of DMR backed by a rigid cavity [7], to achieve perfect absorption with the CMR, the hybrid resonance frequency of one DMR (say DMR-A) under rigid cavity condition with given depth (fixed at 25 mm in this case) must be close to the anti-resonance frequency of the other DMR (say DMR-B), which acts as a rigid wall. The starting trial mass of the central disk of DMR-A for perfect absorption experiments was 120 mg. Figure 1(b) shows the absorption spectra of the CMR when sound waves were incident to the DMR-A side. The absorption did not reach unity at first when the disk mass was 120 mg, because its hybrid resonance frequency was still well above the anti-resonance frequency of DMR-B, and the cavity depth was not well suited. As more putty was added to the disk of DMR-A, the absorption peak moved lower and towards the anti-resonance of DMR-B, while the required cavity depth approached 25 mm. Finally, at 180 mg of disk mass the absorption reached near unity.

The transmission (red curve, left axis), reflection (green curve, left axis), and absorption (purple curve, right axis) spectra of the CMR when sound waves were incident to the DMR-A side are shown in Fig. 1(c). Due to the coupling of the two DMRs via the air cavity, only one main transmission dip is seen around 400 Hz. The smaller feature near 340 Hz is due to the monopole resonance of the CMR. The reflection reaches the first minimum around 200 Hz, while the transmission reaches near unity, due to the first eigenmode of both DMRs. As a result, the absorption is small, and the metasurface becomes highly transparent. At the second reflection minimum of 0.12 around 335 Hz, however, the transmission is also low ($\sim 0.06$), and the metasurface absorbs most (98 %) of the incident waves, even though the area of the membrane is 3.4 ($= (50/27)^2$) times smaller than the impedance tube, and the thickness of the CMR (the length of the hollow cylinder) is less than 1/40 of the corresponding wavelength in air. Indeed, from the surface impedance extracted from the experimental data shown in Fig. 1(d), one can see that near 335 Hz, the imaginary part of the impedance cross the zero line, while the real part is only $\sim 10$ % larger than the air impedance $Z_0$. In the meantime, from the reflection spectrum of the CMR shown in Fig. 1(e) when sound waves were incident to the DMR-B side, one can see that the minimum near 200 Hz due to the first eigenmode is still present, while the one at 335 Hz is completely absent. The surface impedance shown in Fig. 1(f) at the DMR-B side is about $Z = (10 - 32i)Z_0$ at 335 Hz, exhibiting a hard surface that reflects most of the incident sound. The metasurface asymmetric contrast is at the extreme at 335 Hz, i. e., one side is perfectly absorbing and the other side is totally reflecting.

Next, we modified DMR-B to make the metasurface voltage tunable. In Ref. 18, we achieved voltage tunability by turning the central disk into one of the parallel capacitor electrodes. The applied voltage exerted an electrostatic force to the central disk, and the scheme could change the first eigenmode frequency but not the second one, because at the second eigenmode only the portion of the membrane not in direct contact with the central disk was vibrating [3]. Here we use a scheme shown in the insert in Fig. 2(a) that changes the effective free-vibrating area of the membrane not in direct contact with the central disk. The bottom surface of the membrane (purple color) was coated with carbon grease as the soft electrode (golden color). The top surface of the membrane was placed very close to a rigid fishnet ring electrode (golden color) with an outer radius of 27 mm (the same as the membrane) and an inner radius of 20 mm. When DC voltage was applied between the two electrodes, the fishnet ring held the portion of the membrane underneath immobile, even though sound waves could still pass through the fishnet electrode. The size of the membrane that could vibrate in response to the incident sound was reduced from the entire membrane bounded by the rigid frame (light blue) to the portion inside the inner radius of the fishnet ring. Indeed, as the transmission spectra of the modified DMR-B shown in Fig. 2(a) depict, the two transmission peaks and the dip all moved to higher frequency. At 800 V, the first eigenmode shifted by 25 Hz, the anti-resonance frequencies shifted by 55 Hz, and the second eigenmode shifted by 240 Hz as compared to the case without applied voltage. The vibration profile measured by laser vibrometer in Fig. 2(b) shows clearly that the vibration amplitude of the edge portion of the membrane underneath the fishnet ring (marked by the arrows) was held down by the applied voltage. At 800 V, this portion of the membrane was almost motionless, marking the upper limit of the voltage tunability of DMR-B. The incident sound wave intensity was about 120 dB when the laser vibrometer measurements were performed.

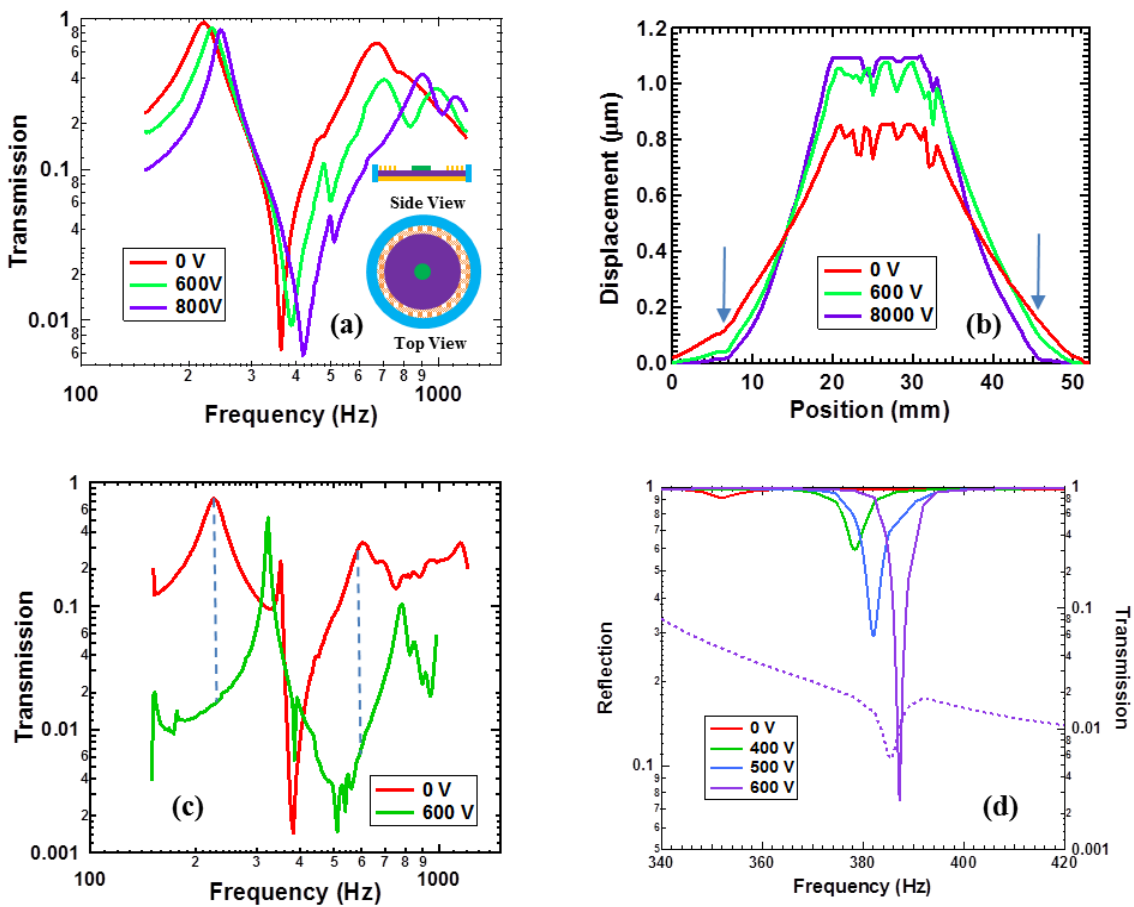

Figure 2. (a) The transmission spectra of DMR-B when measured alone. The insert is a side-view (upper) and a front-view (lower) schematic of the device. (b) The vibration profile of DMR-B at three applied voltages. The arrows mark the position of the inner edge of the fishnet ring electrode. (c) The transmission of the CMR device at 0 and 600 V. The dashed lines mark the frequencies at which the transmission contrast ratio between the two spectra is the largest. (d) The reflection spectra (solid curves) at several applied voltage when sound waves were incident to the DMR-B side. The dashed line is the transmission spectrum of the CMR device at 600 V.

The modified DMR-B was then installed back onto the original CMR. The transmission spectra of the CMR at 0 and 600 V of DC voltage are shown in Fig. 2(c). Large shift in the transmission peak and dip frequencies lead to large transmission contrast. As marked by the two dashed lines at 240 Hz and 590 Hz, the on/off transmission contrast ratio at these two frequencies is over 33 dB. Near these two frequencies, the metasurface functions as a transmission tunable sound switch. In the meantime, the reflection spectra shown in Fig. 2(d) under sound wave incidence onto the DMR-B side change from total reflection, as is in the original metasurface, to perfect absorption, with minimum reflection as low as 0.07 at 388 Hz at 600 V. The on/off contrast ratio in reflection is over 23 dB. In the meantime, as the transmission is 0.006, the absorption is less than 0.005 from unity, indicating 99.5% of incident energy absorption by the metasurface. The DMR-A side remains totally reflecting near 388 Hz. The metasurface is now in the second extreme asymmetric state tuned by the external voltage. At 388 Hz, the DMR-B side is perfectly absorbing and the DMR-A is totally reflecting. Near the original total absorption frequency of 335 Hz at 0 V, the reflection on the DMR-A side is still low but as the transmission is near maximum, leading to low absorption. Therefore, near 335 Hz the CMR is nearly transparent.

Finally, we demonstrate a single-side asymmetric metasurface that exhibits perfect absorption frequency of its front surface tunable by voltage, while the back surface remains totally reflecting throughout the voltage tuning. The basic structure of the metasurface is a hybrid membrane resonator[7] made by a DMR-C backed by a rigid cavity 25 mm in depth. The structure components of the DMR, including the electrode coating on the membrane and the fishnet electrode, are the same as the voltage tunable DMR-B except for the central disk, which is 5 mm in radius and 120 mg in weight. The smaller radius of the central disk in DMR-C, and therefore larger free-vibration membrane area, while having the same disk mass as DMR-B leads to lower eigenmode frequencies than DMR-B. Figure 3 shows the reflection spectra at several applied voltage. The reflection dip frequency changes from 324.2 Hz to 335 Hz for voltage tuning from 0 to 1200 V. The higher saturation voltage observed here as compared to DMR-B is due to the delicate dependence of the tuning function on the detailed structure of the device, such as the flatness of the fishnet electrode and the uniformity of the carbon grease on the bottom surface of the membrane. The reflection minimum is within the range of 0.0205 to 0.15, corresponding to absorption in the deviation range of $4 \times 10^{-4}$ to 0.02 from unity. The reflection contrast between 0 and 600 V at 324.2 Hz is 0.74/0.0205, or 31 dB.

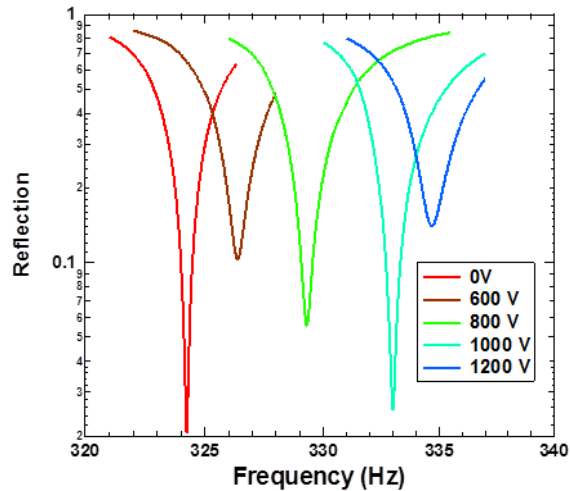

Figure 3. The reflection spectra of the single-side asymmetric metasurface device at several applied voltage.

In summary, we have demonstrated a 3[rd] type of extremely asymmetric acoustic metasurface device made by coupling two decorated membrane resonators with a sealed air column in between. The entire front-view cross section of the device is the same as that of the membrane, as the cavity does not occupy extra lateral space when viewed in the sound propagation direction. This is different from the two earlier types of devices [15, 16], which operate only in waveguides, and the functional components (Helmholtz resonators [15] or hybrid membrane resonators [16]) occupy extra lateral space alongside the waveguides when viewed in the wave propagation direction. As a result, it is very difficult to assemble these devices to form a continuous planar barrier with the entire surface being perfectly absorbing, as the portions of the lateral area occupied by the functional devices are usually highly reflecting. When voltage-tuning components are added to the present device, it becomes a multiple-function metasurface with large tunable transmission and reflection contrast ratio. Unlike the earlier version of voltage tuning devices [18], the present mechanism can change the frequencies of several eigenmodes by changing the effective membrane size. When many of these devices are assembled to form a planar wall, transmission and reflection tunable zone walls with each device individually tunable could perform planar focusing, directional transmission and reflection, and other interesting functions.

Acknowledgement – This work was supported by AoE/P-02/12 from the Research Grant Council of the Hong Kong SAR government.

**References**

1    F. C. Sgard, X. Olny, N. Atalla, and F. Castel, *Appl. Acoust.* **66**, 625 (2005).
2    D. Y. Maa, J. Acoust. Soc. Am. **104**, 2861 (1998).
3    Z. Yang, J. Mei, M. Yang, N. Chan, and P. Sheng, *Phys. Rev. Lett.* **101**, 204301 (2008).
4    Z. Liu, X. Zhang, Y. Mao, Y. Y. Zhu, Z. Yang, C. T. Chan, and P. Sheng, *Science* **289**, 1734 (2000).
5    J. Mei, G. Ma, M. Yang, Z. Yang, W. Wen, and P. Sheng, *Nat. Commun.* **3**, 756 (2012).
6    M. Yang, C. Meng, C. Fu, Y. Li, Z. Yang, and P. Sheng, *Appl. Phys. Lett.* **107**, 104104 (2015).
7    G. Ma, M. Yang, S. Xiao, Z. Yang, and P. Sheng, *Nature Mater.* **13,** 3994 (2014).
8    X. Cai, Q. Guo, G. Hu, and Z. Yang, *Appl. Phys. Lett.* **105**, 121901 (2014).
9    Y. Li and B. M. Assouar, *Appl. Phys. Lett.* **108**, 063502 (2016).
10   X. Wu, C. Fu, X. Li, Y. Meng, Y. Gao, J. Tian, L. Wang, Y. Huang, Z. Yang, and W. Wen, *Appl. Phys. Lett.* **109**, 043501 (2016).
11   Y. D. Chong, L. Ge, H. Cao, and A. D. Stone, *Phys. Rev. Lett.* **105** (2010).
12   W. Wan, Y. Chong, L. Ge, H. Noh, A. D. Stone, and H. Cao, *Science* **331**, 889 (2011).
13   P. Wei, C. Croënne, S. Tak Chu, and J. Li, *Appl. Phys. Lett.* **104**, 121902 (2014).
14   C. Meng, X. Zhang, S. T. Tang, M. Yang, and Z. Yang, *Sci. Rep.* **7**, 43574 (2017)
15   A. Merkel, G. Theocharis, O. Richoux, V. Romero-García, and V. Pagneux, *Applied Physics Letters* **107**, 244102 (2015)
16   Caixing Fu, Xiaonan Zhang, Min Yang, Songwen Xiao, and Z. Yang, *Appl. Phys. Lett.* **110**, 021901 (2017)
17   K. M. Ho, Z. Yang, X. X. Zhang, and P. Sheng, *Appl. Acoust.* **66**, 751 (2005).
18   Songwen Xiao, Guancong Ma, Yong Li, Zhiyu Yang, Ping Sheng, *Appl. Phys. Lett.* **106**, 091904 (2015).